\documentclass{mecbic}
\providecommand{\anno}{2011}
\providecommand{\firstpage}{115}
\providecommand{\lastpage}{116}
\PassOptionsToPackage{colorlinks,citecolor=black,filecolor=black,linkcolor=black,urlcolor=black}{hyperref}
\PassOptionsToPackage{table,colortbl,svgnames}{xcolor}
\usepackage[utf8]{inputenc}
\usepackage[T1]{fontenc}
\usepackage[english]{babel}
\usepackage[makeindex,nomain,acronym,shortcuts]{glossaries} 
\usepackage{cleveref}
\def\eg{{\em e.g.,\,}}

\makeglossaries
\newcommand{\alpina}[0]{AlPiNA}
\crefname{figure}{Figure}{Figures}
\Crefname{figure}{Figure}{Figures}
\crefname{listing}{Listing}{Listings}
\Crefname{listing}{Listing}{Listings}
\crefname{lstlisting}{Listing}{Listings}
\Crefname{lstlisting}{Listing}{Listings}
\def\inst{University of Geneva}
\title{Modelling of Genetic Regulatory Mechanisms with GReg}
\author{Nicolas Sedlmajer \institute{\inst} \email{sedlmaj0@etu.unige.ch}
\and Didier Buchs \institute{\inst} \email{didier.buchs@unige.ch}
\and Steve Hostettler \institute{\inst} \email{steve.hostettler@unige.ch}
\and Alban Linard \institute{\inst} \email{alban.linard@unige.ch}
\and Edmundo Lopez \institute{\inst} \email{edmundo.lopez@unige.ch}
\and Alexis Marechal \institute{\inst} \email{alexis.marechal@unige.ch}
}
\def\titlerunning{Modelling of Genetic Regulatory Mechanisms with GReg}
\def\authorrunning{N. Sedlmajer,  D. Buchs, S. Hostettler\\\ A. Linard, E. Lopez and A. Marechal}
\begin{document}
\maketitle
\pagestyle{plain}
\pagenumbering{arabic}
\setcounter{page}{115}
%
\begin{abstract}
Most available tools propose simulation frameworks to study models of biological systems,
but simulation only explores a few of the most probable behaviours of the system.
On the contrary, techniques such as model checking, coming from IT-systems analysis,
explore all the possible behaviours of the modelled systems, thus helping to identify emergent properties.
A main drawback from most model checking tools in the life sciences domain is that they take as input a language designed for computer scientists, that is not easily understood by non-expert users.
We propose in this article an approach based on DSL.
It provides a comprehensible language to describe the system while allowing the use of complex and powerful underlying model checking techniques.
\end{abstract}
\medskip

However, investigation through formal models of biological systems is not as widely spread as in other natural sciences such as chemistry and physics.
This is partly due to the complexity of the living systems themselves, the strenuosity to formalize biological concepts,
the difficulty in performing experiments on living systems in order to test a model, {\it etc}.
Another reason is that formal models require the use of formalisms (\eg{} Petri Nets (PNs)), which are usually too complex for non-experts.
To circumvent this difficulty, we propose to use the Domain Specific Languages (DSL) approach.
A DSL is a language designed to be understandable by a domain expert and at the same time translatable into a formal language.
This paper presents Gene Regulation Language (GReg), a DSL for the modelling of genetic regulatory mechanisms through the regulatory network approach~\cite{Chaouiya11,Thomas91}.
\smallskip

The main idea of regulatory networks is to model inter-biological reactions through a set of interdependent biological rules.
This can be seen as a set of discrete modules having strong interconnections.
The occurrence of interesting events in the biological system can be represented as logical properties expressed on the states of these modules.
This is very similar to the kind of properties computer scientists validate on hardware and software systems (deadlocks, invariants, reachability, \ldots).
\smallskip

Among the tools available in this domain, the main analysis approach for regulatory networks is {\it simulation}.
Simulation is generating and analyzing a limited sample of possible system behaviours.
This technique is not convenient when the main purpose of the research is to look for rare or abnormal behaviours (\eg{}~cancer).
The main approach in this case is to use {\it model checking} instead of simulation.
Model checking consists in generating and analysing all the possible states of the system.
Naturally, this technique suffers from the drawback of the enormous number of possible states of biological systems.
\smallskip

It is interesting to note that this problem is well-known to the model checking community in computer science,
where it is called the {\it state space explosion}.
There is a parallel between cellular interactions and software systems in that the state space explosion is partly due to their concurrent nature,
but mainly due to the number of molecular species, and their respective populations, that are present in living organisms.
Therefore, we can apply techniques that have been developed for the model checking of hardware and software systems to biological interactions.
Approaches based on a symbolic encoding of the state space~\cite{McMillan92} are particularly well-suited for this.
\medskip

In this paper we show a work in progress in our group.
We present  GReg, a DSL dedicated to the modelling of genetic regulatory mechanisms.
GReg is given as an example of the DSL-based verification process.
We describe the creation of a language tailored to the understanding of domain experts,
and how this language can be translated into a formal model where model checking can be applied.
Model checking is a very well known verification technique used in computer science.
Its main advantage is the complete exploration of the state space of the model,
thus allowing to discover rare but potentially interesting events.
A query language to express the properties of such events is embedded with GReg.
\medskip

Three axes of development lie ahead:
improving the expressivity of the modelling and query languages,
assessing the usability of the approach and
exploring the mitigation of the state space explosion.
\vskip 1em

\begin{description}
\item[Extending the expressivity of GReg]
Concepts such as time and probabilities play an important role in biology and are therefore good candidates for a language extension.
As Algebraic Petri Nets (APNs), the current underlying formalism, do not support these notions, such extension would require to change the target platform.
Good examples of target formalisms are timed Petri nets and stochastic Petri nets.
The techniques used to create GReg allow changing the target language without changing the language itself.
\item[Improving the usability of our tool.]
Although highly efficient, textual languages are usually not as intuitive as graphical languages.
On the other hand, graphical domain specific languages are especially good in the early phase of the modeling as well as for documentation,
but they are often less practical when the model grows.
The tools in Eclipse Modelling Project (EMP) that were used to create GReg allow us to define a graphical version of the same language, thus keeping the best of both worlds.
Another way to ease the modelling phase is to allow import/export of models from/to other formalisms and standards such as Systems Biology Markup Language (SBML) and to integrate GReg with {\it Cytoscape} framework through its plug-in mechanism.
\item[Mitigating the state space explosion.]
So far, we have done little experimentation in this area for real biological processes.
This weakness is being worked on as we speak using studies found in the literature and adapted to GReg.
We are also working on a more detailed comparison to other tools dedicated to biological problems, \eg GinSim.
Nevertheless, we conducted several studies on usual IT protocols and software models that show that \alpina{} can handle huge state spaces~\cite{buchs:petrinets:2010}.
This suggests promising results in the regulatory mechanisms domain.
\end{description}
\bibliographystyle{plain}

\end{document}